\documentclass[showpacs,preprintnumbers,prd,nofootinbib,floats,amssymb,floatfix]{revtex4}
\usepackage{graphicx}
\usepackage{amsxtra}
\usepackage{amssymb}
\usepackage{amsmath}
\usepackage{bbold}
\usepackage{physics}
\usepackage{bm}
\usepackage{tensor}
\usepackage{xcolor}
\usepackage{graphicx}
\graphicspath{ {Imagenes/} }
\usepackage{hyperref}
\usepackage{amssymb}
\usepackage{amstext}
\usepackage{amsmath}
\usepackage{cleveref}
\usepackage{stackrel}
\usepackage{blkarray}
\begin{document}
\title{Canonical description of Pontryagin and Euler classes with  a Barbero-Immirzi parameter }
 \author{Alberto Escalante}  \email{aescalan@ifuap.buap.mx}
 \author{Edmundo Su\'arez-Polo} \email{esuarez@ifuap.buap.mx} 
\affiliation{Instituto de F\'isica, Benem\'erita Universidad Aut\'onoma de Puebla. \\ Apartado Postal J-48 72570, Puebla Pue., M\'exico,}
\author{Luis A. Huerta-del Campo}\email{luis.huertadelc@alumno.buap.mx}
\affiliation{Facultad de Ciencias Físico Matemáticas, Benemérita Universidad Autónoma de Puebla, Apartado postal
1152, 72001 Puebla, Pue., Mexico}
 
\begin{abstract}
A detailed canonical analysis for Pontryagin and Euler classes with a Barbero-Immirzi [BI] parameter is developed. We rewrite the topological invariants by introducing a set of Holst-like variables, and then study the set of all constraints. We report the complete canonical structure and the symmetries of the theory; we count the physical degrees of freedom and identify reducibility conditions among the constraints. In addition, in our results, if we consider the $BI$ parameter takes the value of $\gamma = \pm i $, then the self-dual representation of these invariants is reproduced. Finally, we couple the invariants to the Holst action and explore the canonical analysis.

\end{abstract}
 \date{\today}
\preprint{}
\maketitle

\section{Introduction}

It is well known that topological theories play an important role in physics. In fact, if they are viewed as field theories, they provide an interesting connection between geometry and General Relativity (GR). From a classical standpoint, they are devoid of physical degrees of freedom and covariant under diffeomorphisms. These important symmetries provide excellent testing grounds for ideas about constructing a background-independent quantum gravity theory \cite{1}. Moreover, topological theories play a significant role in the self-dual canonical approach to $GR$; notably, when $GR$ incorporates topological terms such as the Pontryagin and  Euler, these invariants, while not affecting the equations of motion, contribute significantly to the canonical structure of the theory; these terms act as generating functionals of canonical transformations, which may have physical relevance in quantum gravity  \cite{2}. On the other hand, in the context of classical field theory, either the Euler or the Pontryagin classes serve as essential components in constructing the noncommutative form of topological gravity. Moreover, those invariants have been important in the study of noncommutative gravity, which is treated like noncommutative Yang-Mills theory \cite{3}. In this regard, if the gauge group of the action is $SO(3,1)$, the study of the topological invariants is performed by introducing the self-dual (anti-self-dual) formulation (from now on, we will only refer to the self-dual terminology). In this scenario, the fundamental fields of the theory are complex; the noncommutative action is expressed in terms of an $SL(2,C)$ invariant action. It is worth noting that several works use the self-dual formulation of these invariants, which simplifies their study from either a Lagrangian or a canonical perspective. On the other hand, from a mathematical point of view, the final expressions exhibit symmetries that are not evident from the standard viewpoint of the fundamental variables \cite{4}.  \\
Furthermore, in $GR$,  from the canonical point of view,  the use of the self-dual variables facilitates the canonical study \cite{5, 6, 7, 8, 9, 10}. In fact, the use of these variables leads to an important simplification of the theory equations. In this framework, both the constraints and the evolution equations of the canonical general relativity become simple polynomials of the field variables. Nevertheless, the price to pay for these simplifications is that the self-dual (so-called Ashtekar) variables are complex, and therefore, the Ashtekar canonical formulation describes complex general relativity. In order to obtain the real physical degrees of freedom, one needs to append a posteriori appropriate reality conditions \cite{11, 12}. In this regard, the study of $GR$ and the addition of topological terms proceed in the same way; for example, the introduction of self-dual variables facilitates their study \cite{13}. Nonetheless, it is well known that there is an alternative to the self-dual framework, the so-called Holst variables. In fact, in $GR$, the Holst variables involve an arbitrary parameter, the so-called Barbero-Immirzi $[BI]$ parameter (which we will also call $\gamma$) \cite{13a}. In this description, if $\gamma=i$ then the self-dual formulation of $GR$ is reproduced. On the other hand, if $\gamma=1$, the Barbero formulation is recovered \cite{14}. In this regard, a general study of the $PE$ invariants and $ GR$ plus the $PE$ invariants is mandatory; for example, a general study with an arbitrary $BI$ parameter is necessary to connect the self-dual formulation with real Barbero's variables \cite{14a}. The physical meaning of the $ BI $ remains an open question, encouraging further research in this area \cite{14b}. In this respect, we propose to analyze the $PE$ invariants in terms of an arbitrary $BI$ parameter. In our formulation, if $\gamma = \pm i$, we obtain the self-dual description of $PE$, and if $\gamma = 1$, we shall obtain the Barbero formulation of these invariants. Thus, we expect to contribute to the debate on the meaning of the $BI$ parameter, for instance, in our results with respect to the $PE$ invariants;  if $\gamma$ is real, then a new Gauss-like constraint occurs in the canonical description, the reducibility conditions between the constraints remain, and the counting of physical degrees of freedom yields zero. When the $PE$ are added to the Holst action, we will observe that for arbitrary  $\gamma$  there is a contribution of the $BI$ parameter in the constraints, also in the algebra between them. Furthermore, if $\gamma$ is real, there exists a new canonical description of the theory, the constraints are not polynomial, but the algebra between them is closed. \\
The paper is organized as follows. In section II, by introducing a Holst-like variable, the $PE$ invariants are expressed in terms of an arbitrary $\gamma$ parameter. Then the canonical study is performed. In section III, the results of previous sections are considered in Holst's action. We provide a complete canonical description of the theory and analyze the cases in which $\gamma$ is arbitrary. Section IV is devoted to conclusions. 

\section{ The  Pontryagin and  Euler classes with a Barbero-Immirzi parameter} \label{ep-bi}
In this section we will rewrite the topological invariants with a $BI$ parameter \cite{14, 15, 16, 17}. We start with the following action  
\begin{equation}
\label{a1}
 S[\omega] = \int \star R_{IJ}[\omega] \wedge R^{IJ}[\omega] + \frac{1}{\gamma} \int R_{IJ}[\omega] \wedge R^{IJ}[\omega],
\end{equation}
where the first term is identified with the Euler class and the second one is the Pontryagin invariant. Here, $I, I, K,..=0, 1, 2, 3$, $\omega$ is the Lorentz connexion, the star product is defined by    $\star T^{IJ} = \frac{1}{2} \epsilon^{IJKL} T_{KL}$  where $ \epsilon^{IJKL}$ is the volume 4-form associated with the internal metric $\eta_{IJ}=(-1, 0, 0, 0)$, $\epsilon^{0123}=1$,   $R_{IJ}[\omega]= d\omega_{IJ} + \omega_{I}{^{K}}\wedge \omega_{KJ}$ is the Lorentz curvature and $\gamma$ is the so-called  BI parameter. By introducing the following field $F^{IJ}= \star R^{IJ} + \frac{1}{\gamma} R^{IJ}$,  we can write the  action (\ref{a1}) in  the following form
\begin{equation}
\label{pe2}
S[\omega] = -\frac{\gamma^2}{\gamma^2 + 1}\int ( \star F_{IJ} - \frac{1}{\gamma} F_{IJ}) \wedge F^{IJ}.
\end{equation}
To facilitate the computations, we introduce the following field $\tilde{F}^{IJ} = \star F^{IJ} - \frac{1}{\gamma} F^{IJ}$, then the action will take a new fashion
\begin{equation}
\label{pe3}
    S[\omega] = -\frac{\gamma^2}{\gamma^2 + 1}\int \tilde{F}^{IJ} \wedge F^{IJ},
\end{equation}
where, the curvature and the fields are given by 
\begin{equation}
R^{IJ}_{\mu \nu} = \partial_\mu \omega_\nu^{IJ} - \partial_\nu \omega_\mu^{IJ} + \omega_\mu^{IK}\omega_{\nu \, K}^{\, \, \, J} - \omega_\mu^{JK}\omega_{\nu \, K}^{\, \, \, I},
\end{equation}
\begin{equation}
F_{\mu \nu}^{IJ} = \frac{1}{2}\epsilon^{IJKL} R_{\mu \nu \, KL} + \frac{1}{\gamma} R_{\mu \nu}^{IJ}, 
\end{equation}
\begin{equation}
\label{tilF}
\tilde{F}_{\mu \nu}^{IJ} = \frac{1}{2}\epsilon^{IJKL} F_{\mu \nu \, KL} - \frac{1}{\gamma} F_{\mu \nu}^{IJ}.
\end{equation}
By performing the 3+1 decomposition,  the action (\ref{pe3}) takes the form 
\begin{equation}
\label{Sg}
 S[\omega] = -\frac{\gamma^2}{\gamma^2 + 1} \int d^4 x \, \eta^{abc}(2 \tilde{F}_{b c \, 0 i} F_{0 a}^{\, \, \, 0 i} + \tilde{F}_{b c \, i j} F_{0 a}^{\, \, \, i j}), 
\end{equation} 
where we define  $ \eta^{abc} \equiv \epsilon^{0abc}$ and $\epsilon^{ijk} \equiv \epsilon^{0ijk}$, $i, j, k,..=1, 2, 3$. Now, after long computations we find that  the fields  $\tilde{F}_{b c \, 0 i} $ and  $\tilde{F}_{b c \, i j} $ given in (\ref{tilF}), can be written as 
\begin{equation}
\label{eq: ep1}
\begin{split}
\tilde{F}_{b c \, 0 i} =& -\frac{\gamma^2 + 1}{\gamma^2} R_{b c \, 0i}, \\
\tilde{F}_{b c \, i j} =& -\frac{\gamma^2 + 1}{\gamma^2} R_{b c \, i j}.
\end{split}
\end{equation}
At this point, we will not use either the Holst variables or the variables reported in \cite{2}; instead, we will introduce the following variables   
\begin{equation}
\label{cv1}
    \omega_{\alpha}^{\, i j} = \epsilon^{i j}_{\, \, k} \omega_{\alpha}^{\; k},
\end{equation}
\begin{equation}
\label{cv2}
        \omega_{\alpha}^{\; k} = \frac{1}{\gamma}(A_{\alpha}^{\; k} - \omega_{\alpha}^{\; 0 k}),
\end{equation}
in terms of these variables, we find that the curvature $R^{IJ}$ and the field  $F^{IJ}$ are expressed as  
\begin{equation}
\label{eq: ep2}
\begin{split}
R_{\mu \nu }^{0i} =& \partial_\mu \omega_{\nu}^{\; 0 i} - \partial_\nu \omega_{\mu}^{\; 0 i} - \frac{2}{\gamma} \epsilon^{i}_{\, j k} \omega_{\mu}^{\; 0 j} (A_{\nu}^{\; k} - \omega_{\nu}^{\; 0 k}), \\
R_{\mu \nu }^{ij} =& \epsilon^{i j}_{\, \, k} \Bigg [ \partial_\mu A_{\nu}^{\;  k} - \partial_\nu A_{\mu}^{\;  k} - \epsilon^{k}_{\, m n} A_{\mu}^{\; m} A_{\nu}^{\; n} + \frac{1}{\gamma^2} \epsilon^{k}_{\, m n} (A_{\mu}^{\; m} \omega_{\nu}^{\; 0 n} - A_{\nu}^{\; m} \omega_{\mu}^{\; 0 n}) \\
&\quad\quad -\frac{1}{\gamma}(\partial_\mu \omega_{\nu}^{\; 0 k} - \partial_\nu \omega_{\mu}^{\; 0 k}) + \frac{\gamma^2 - 1}{\gamma^2} \epsilon^{k}_{\, m n} \omega_{\mu}^{\; 0 m} \omega_{\nu}^{\; 0 n} \Bigg ], \\
F_{\mu \nu}^{0 i} =& \frac{1}{\gamma}(\partial_\mu A_{\nu}^{\;  i} - \partial_\nu A_{\mu}^{\;  i}) - \frac{\epsilon^{i}_{\, j k}}{\gamma^2} A_{\mu}^{\; j} A_{\nu}^{\; k} + \frac{\gamma^2 + 1}{\gamma^2} \epsilon^{i}_{\, j k} \omega_{\mu}^{\; 0 j} \omega_{\nu}^{\; 0 k}, \\
F_{\mu \nu}^{i j} =& \epsilon^{i j}_{\, \, k} \Bigg [ \frac{1}{\gamma^2}(\partial_\mu A_{\nu}^{\;  k} - \partial_\nu A_{\mu}^{\; k}) - \frac{1}{\gamma^3} \epsilon^{k}_{\, m n} A_{\mu}^{\; m} A_{\nu}^{\; n} + \frac{\gamma^2 + 1}{\gamma^3} \epsilon^{k}_{\, m n} (A_{\mu}^{\; m} \omega_{\nu}^{\; 0 n} - A_{\nu}^{\; m} \omega_{\mu}^{\; 0 n})\\
&\quad\quad - \frac{\gamma^2 + 1}{\gamma^2}(\partial_\mu \omega_{\nu}^{\; 0k} - \partial_\nu \omega_{\mu}^{\; 0 k}) - \frac{\gamma^2 + 1}{\gamma^3}\epsilon^{k}_{\, m n} \omega_{\mu}^{\; 0 m} \omega_{\nu}^{\; 0 n} \Bigg ].
\end{split}
\end{equation}
Furthermore, by using   (\ref{eq: ep1} - \ref{eq: ep2}), it is possible to show that the action  (\ref{Sg}) can be written as
\begin{equation}
\label{eq: S1}
\begin{split}
S = \int d^4 x \,\eta^{abc}&(2 R_{b c \, 0i} F_{0 a}^{\, \, 0 i} + R_{b c \, i j} F_{0 a}^{\, \, i j}) \\
= \int d^4 x \,\eta^{abc}& \Bigg \{ 2 R_{b c \, 0i} \Bigg (\frac{1}{\gamma}(\dot{A}_a^{\;  i} - \partial_a A_0^{\;  i}) - \epsilon^{i}_{\, j k} A_0^{\; j} A_a^{\; k} + \frac{\gamma^2 + 1}{\gamma^2} \epsilon^{i}_{\, j k} \omega_0^{\; 0 j} \omega_a^{\; 0 k} \Bigg ) \\
&+ \epsilon^{i j}{_{\, \, k}}R_{b c \, i j} \Bigg [ \frac{1}{\gamma^2}(\dot{A}_{a}^{\; k} - \partial_a A_{0}^{\;  k}) - \frac{1}{\gamma^3} \epsilon^{k}_{\, m n} A_{0}^{\; m} A_{a}^{\; n} + \frac{\gamma^2 + 1}{\gamma^2} \epsilon^{k}_{\, m n} (A_{0}^{\; m} \omega_{a}^{\; 0 n} - A_{a}^{\; m} \omega_{0}^{\; 0 n}) \\
&\quad\quad + \frac{\gamma^2 + 1}{\gamma^2}(\dot{\omega}_{a}^{\; 0 k} - \partial_a \omega_{0}^{\; 0 k}) - \frac{\gamma^2 + 1}{\gamma^3}\epsilon^{k}_{\, m n} \omega_{0}^{\; 0 m} \omega_{a}^{\; 0 n} \Bigg ] \Bigg \}, 
\end{split}
\end{equation}
where we can identify the following momenta
\begin{equation}
\label{momenta}
    \pi^{a}_{\; k} = \eta^{abc}\left(\frac{2}{\gamma} R_{b c \, 0 k} + \frac{1}{\gamma^2}\epsilon^{i j}_{\, \, k} R_{b c \, i j}\right),
\end{equation}
\begin{equation}
\label{momenta2}
    \tilde{\pi}^{a}_{\; k} = -\left(\frac{\gamma^2 + 1}{\gamma^2}\right)\eta^{abc}\epsilon^{i j}_{\, \, k} R_{b c \, i j},
\end{equation}
these are canonically conjugate to  $A_{a}^{\; i}$ and  $\omega_{a}^{\; 0 i}$,  respectively. An explicit form of the momenta in terms of the variables (\ref{cv2}) is given by 
\begin{eqnarray}
\label{momenta}
 \pi^{a}_{\; i} &=& 2\eta^{abc}\Big( - \Big( \frac{\gamma^2 + 1}{\gamma^3}\Big) \big( \partial _b \omega^{0i}_c -\partial _c \omega^{0i}_b \big)  + \frac{1}{\gamma^3} \big(\partial _b A{^{i}}_c - \partial _c A{^{i}}_b  \big) -\frac{\epsilon_{ijk}}{ \gamma^4} A^j_bA^k_c \nonumber \\
 & -&  \Big( \frac{\gamma^2 + 1}{\gamma^4}\Big)  \epsilon_{ikl} \omega^{0k}_{b} \omega_{c}{^{0l}} + \frac{2 (\gamma^2 +1)}{\gamma^4} \epsilon_{ijk} A_b ^j\omega^{0k}_{c}     \Big), \nonumber\\
  \tilde{\pi}^{a}_{\; i}&=& -  \eta^{abc} \Big(\frac{\gamma^2 + 1}{\gamma^2}\Big) \Big\{  \frac{4}{\gamma} \big( \partial _b A_{ci} -\partial _b\omega^{0i}_c \big) + \frac{2 (\gamma^2 - 1)}{\gamma^2}\epsilon_{ikl}\omega_b^{0k}\omega_c^{0l} -\frac{2}{\gamma^2} \epsilon_{ijk}A^j_bA^k_c +\frac{4}{\gamma^2}\epsilon_{ikl}A^k_b \omega_c^{0l} \Big\}. \nonumber \\
\end{eqnarray}
With the canonical variables identified, the fundamental Poisson brackets will be given by 
\begin{equation}
\begin{split}
\{ A_{a}^{\; i} (\vec x),  \pi^{b}_{\; j} (\vec y) \} &= \delta_j^i \delta_{a}^{b} \delta^3 (\vec x - \vec y), \\
\{ \omega_{a}^{\; 0 i} (\vec x),  \tilde{\pi}^{b}_{\; j} (\vec y) \} &= \delta_j^i \delta_a^b \delta^3 (\vec x - \vec y).
\end{split}
\end{equation}
Thus, the introduction of the momenta (\ref{momenta}) and  (\ref{momenta2}) into (\ref{eq: S1}), the action takes the following canonical form
\begin{equation}
\label{AcB}
    \begin{split}
        S = \int d^4 x \Bigg \{ \, &\pi^{a}_{\; k} \dot{A}_{a}^{\, k} + A_0^{\, i} \left( \partial_a \pi^{a}_{\; i} - \frac{1}{\gamma}\epsilon_{i j}^{\, \, k}(A_{a}^{\, j} \pi^{a}_{\; k} + \omega_{a}^{\, 0 j} \tilde{\pi}^{a}_{\; k}) \right) \\
        & + \tilde{\pi}^{a}_{\; k} \dot{\omega}_{a}^{\, 0 k} + \omega_{0}^{\, 0 i} \left( \partial_a \tilde{\pi}^{a}_{\; i} + \frac{1}{\gamma} \epsilon_{i j}{^{k}} (\omega_{a}^{\, 0 j} (2 \tilde{\pi}^{a}_{\; k} + (\gamma^2 + 1) \pi^{a}_{\; k}) - A_{a}^{\, j}\tilde{\pi}^{a}_{\; k}) \right)\Bigg \}.
    \end{split}
\end{equation}
It is worth commenting if the  $BI$ parameter takes the value $\gamma = \pm i$, the canonical momenta  $\tilde{\pi}^{a}_{\; k}$ defined by  (\ref{momenta2}) vanish, thus, in the self-dual scenario there is not contribution of $\tilde{\pi}^{a}_{\; k}$ to the action (\ref{AcB}). With this particular value of the $BI$ parameter,  our results reproduce the self-dual formulation of the $PE$ invariants reported in  \cite{13}, say 
\begin{equation}
    \begin{split}
        S = \int d^4 x \Bigg \{ \, &\pi^{a}_{\; k} \dot{A}_{a}^{\, k} + A_0^{\, i} \left( \partial_a \pi^{a}_{\; i} - \frac{1}{i}\epsilon_{i j}{^{k}}A_{a}^{\, j} \pi^{a}_{\; k} \right)  \Bigg \}, 
    \end{split}
\end{equation}
where $ \partial_a \pi^{a}_{\; i} - \frac{1}{i}\epsilon_{i j}{^{k}}A_{a}^{\, j} \pi^{a}_{\; k}$ ist the so-called Gauss constraint. Moreover, we can observe that there is not a Hamiltonian constraint, because the invariants are topological.\\
On the other hand, from (\ref{AcB}) with  arbitrary $\gamma$, we identify the following constraints of the theory
\begin{equation}
\label{const1}
    C^{a}_{\; k} \equiv \pi^{a}_{\; k} - \eta^{abc} \left( \frac{2}{\gamma}R_{b c \, 0 k} + \frac{1}{\gamma^2}\epsilon^{i j}{_{ k}}R_{b c \, i j} \right) \approx 0,
\end{equation}
\begin{equation}
    \tilde{C}^{a}_{\; k} \equiv  \tilde{\pi}^{a}_{\; k} - \frac{\gamma^2 + 1}{\gamma^2} \eta^{abc}\epsilon^{i j}{_{k}} R_{b c \, i j} \approx 0,
\end{equation}

\begin{equation}
\label{Gauss}
    G_i \equiv \partial_a \pi^{a}_{\; i} - \frac{1}{\gamma} \epsilon_{i j}{^{ k}}(A_{a}^{\, j} \pi^{a}_{\; k} + \omega_{a}^{\, 0 j} \tilde{\pi}^{a}_{\; k}) \approx 0,
\end{equation}
\begin{equation}
    \tilde{G}_k \equiv \partial_a \tilde{\pi}^{a}_{\; k} + \frac{1}{\gamma} \epsilon_{i j}{^{k}}[ \omega_{a}^{\, 0 j} (2 \tilde{\pi}^{a}_{\; k} + (\gamma^2 + 1)\pi^{a}_{\; k}) - A_{a}^{\, j} \tilde{\pi}^{a}_{\; k}] \approx 0.
\end{equation}
Those constraints satisfy  the following algebra  
\begin{equation}
    \{ C^{a}_{\; k}(\vec x), C^{b}_{\; l}(\vec y) \} = 0,
\end{equation}
\begin{equation}
    \{ \tilde{C}^{a}_{\; k}(\vec x), \tilde{C}^{b}_{\; l}(\vec y) \} = 0,
\end{equation}
\begin{equation}
    \{ C^{a}_{\; k}(\vec x), \tilde{C}^{b}_{\; l}(\vec y) \} = 0,
\end{equation}
\begin{equation}
    \{ G_i (\vec x), G_j (\vec y) \} = -\frac{1}{\gamma}\epsilon_{i j}{^ {k}}G_k (\vec x)\delta^3 (\vec x - \vec y),
\end{equation}
\begin{equation}
    \{ \tilde{G}_i (\vec x), \tilde{G}_j (\vec y) \} = \epsilon_{i j}{^{ k}} \left[ \frac{\gamma^2 + 1}{\gamma}G_k (\vec x) + \frac{2}{\gamma}\tilde{G}_k (\vec x) \right] \delta^3 (\vec x - \vec y),
\end{equation}
\begin{equation}
    \{ G_i (\vec x), \tilde{G}_j (\vec y) \} = -\frac{1}{\gamma}\epsilon_{i j}{^{ k}}\tilde{G}_k (\vec x)\delta^3 (\vec x - \vec y),
\end{equation}
\begin{equation}
    \{ C^{a}_{\; i}(\vec x), G_j (\vec y) \} = -\frac{1}{\gamma}\epsilon_{i j}{^{k}} C^{a}_{\; k}(\vec x)\delta^3 (\vec x - \vec y),
\end{equation}
\begin{equation}
    \{ \tilde{C}^{a}_{\; i}(\vec x), G_j (\vec y) \} = -\frac{1}{\gamma}\epsilon_{i j}{^{ k}}\tilde{C}^{a}_{\; k}(\vec x)\delta^3 (\vec x - \vec y),
\end{equation}
\begin{equation}
    \{ C^{a}_{\; i}(\vec y), \tilde{G}_j (\vec y) \} = -\frac{1}{\gamma}\epsilon_{i j}{^{k}}\tilde{C}^{a} _{\; k}(\vec x)\delta^3 (\vec x - \vec y),
\end{equation}
\begin{equation}
\label{const2}
    \{ \tilde{C}^{a}_{\;_i}(\vec y), \tilde{G}_j (\vec y) \} = \epsilon_{i j}{^{k}} \left[ \frac{\gamma^2 + 1}{\gamma} C^{a}_{\; k}(\vec x) + \frac{2}{\gamma} \tilde{C}^{a}_{\; k}(\vec x) \right] \delta^3 (\vec x - \vec y).
\end{equation}
We observe that the algebra is closed and therefore the constraints are first class. It is worth noting that the $BI$  parameter occurs in the constraint algebra, and this fact could be important in the quantization program. On the other hand, we also observe that if the $BI$ parameter takes the value  $\gamma = \pm i$, then the constraint algebra is reduced to the algebra of the self-dual $PE$ formulation. Moreover, if $\gamma= 1$, the momenta $\tilde{\pi}^a_i$ do not vanish, and we must obtain the Barbero formulation of the $PE$ theory, this result extends those reported in the literature. In fact, the constraints are not polynomial anymore, but the algebra between the constraints remains. Furthermore, we notice that  $\partial_a C^{a}_{\; i} = G_i$ and  $\partial_a \tilde{C}^{a}_{\; i} = \tilde{G}_i$, this is because the Bianchi identities, hence, this implies that the constraints $ C^{a}_{\; i}$ and $\tilde{C}^{a}_{\; i}$ are reducible; they are not independent \cite{17}. In fact, reducibility conditions in the constraints are usual in this kind of theory, where the equations of motion are the Bianchi identities. By taking into account the reducibility conditions, there are 36 canonical variables, and   there are 18 independent first-class constraints; therefore, the counting of physical degrees of freedom yields

\begin{equation}
\mbox{D.o.F.} = \frac{1}{2}(36 - 2(18)) = 0,
\end{equation}
this result was expected because we are dealing with topological theories \cite{17}. \\
On the other hand, let us see the following. Again,  if $\gamma = i$, then the momenta (\ref{momenta}) are reduced to
\begin{equation}
\pi^a_i=-2i\eta^{abc}F_{bci}, 
\end{equation}
where $F_{bci}=\partial_bA_c^i-\partial_c A_b^i - \epsilon_{ijk}A^j_bA^k_c $, and the Gauss constraint  is reduced to
\begin{equation}
G_i=\partial_a\pi^a_i+i \epsilon{_{ij}}^kA^j_a\pi^a_k\approx 0, 
\end{equation}
we can observe that the Gauss constraint has not the standard form reported in the literature, however, if we perform the change of  $ A_a^{\, i} \rightarrow  - \gamma  \tilde{A}_a^{\, i} $ into all our calculations,   and taking $\gamma=i$, all our results are reduced to the desired form, say, the Gauss constraint will take the form  $G_i=\partial_a\pi^a_i+\epsilon{_{ij}}^k \tilde{A}^j_a\pi^a_k\approx 0$, as expected,  where the variables $\tilde A$'s are the so-called Ashtekar variables. 
\section{Holst action plus the Euler and Pontryagin invariants with a Barbero-Immirzi parameter}
Now, we will couple the $PE$ invariants to the Holst action. To compare the new results with those reported in the literature, we will use a different set of canonical variables. In fact, in previous sections, we used the canonical variables $\omega^{0i}_\alpha$; however, in the coupling to gravity, we will use the standard ones used in the literature. We start with the following action 
\begin{equation}
S[e, \omega]= \int dx^4 \epsilon^{\mu \nu \rho \sigma} \Big(  \frac{1}{4} \epsilon_{IJKL} e^I_\mu e^J_\nu R^{KL}_{\rho \sigma}  +  \frac{1}{2\gamma} e_{I \mu}e_{\nu J}R^{IJ}_{\rho \sigma} +\frac{1}{4} R^{IJ}_{\mu \nu }R_{IJ \rho \sigma} -\frac{1}{4 \gamma}
\epsilon_{IJKL}{R}^{IJ}_{\mu \nu }R^{KL}_{ \rho \sigma } \Big),
\end{equation}
where $e_{\alpha}^I$ is the tetrad and the curvature has been defined above. By performing the $3+1$ decomposition, then by taking into account the so-called time gauge, say $e^0_a=0$, and  now using the change of variable $\omega^{0i}_\alpha= A^i_{\alpha}- \frac{\omega^i_\alpha}{\gamma} $, the action, after long computations, acquires the following canonical form
\begin{align}
\label{Holst}
		\begin{split}
			S 
			&= \int dx^4 \, \Biggl\{  \frac{\Xi_{i}^{a}} {\gamma} \dot{\tilde{{A}}}^i_a+ \tilde{\pi}_{i}^{a} \dot{\omega}^a_i  -\frac{\tilde{N}}{\gamma^2}\tensor{\epsilon}{_i^j^k}\left(\Xi_{j}^{a}-\pi_{j}^{a}\right)\left(\Xi_{k}^{b}-\pi_{k}^{b}\right)\left[F_{ab}^{i} + (\gamma^2+1)R_{ab}^{i}\right]\\
			&- \frac{N^{b}}{\gamma}\Xi_{i}^{a}F_{ab}^{i} + \left[\frac{A_{0}^{i}}{\gamma^2} + \left(\frac{\gamma^2+1}{\gamma^3}\right)\omega_{0}^{i}\right]\left(\partial_{a}\Xi_{i}^{a} + \tensor{\epsilon}{_i_j^k}\tilde{A}_{a}^{j}\Xi_{k}^{a}\right)\\
			&+ \left[\left(\frac{\gamma^2+1}{\gamma^2}\right)A_{0}^{i} - \left(\frac{\gamma^2+1}{\gamma^3}\right)\omega_{0}^{i}\right]\left(\partial_{a}\Xi_{i}^{a} - \tensor{\epsilon}{_i_j^k}\omega_{a}^{j}\Xi_{k}^{a}\right)\\
			&+ \left(-\frac{A_{0}^{i}}{\gamma} + \frac{\omega_{0}^{i}}{\gamma^2}\right)\left(\partial_{a}\tilde{\pi}_{i}^{a} + \tensor{\epsilon}{_i_j^k}\tilde{A}_{a}^{j}\tilde{\pi}_{k}^{a}\right)\\
			&+ \left[\frac{A_{0}^{i}}{\gamma} - \left(\frac{1-\gamma^2}{\gamma^2}\right)\omega_{0}^{i}\right]\left(\partial_{a}\tilde{\pi}_{i}^{a} - \tensor{\epsilon}{_i_j^k}\omega_{a}^{j}\tilde{\pi}_{k}^{a}\right) \Biggr\}
		\end{split}
	\end{align}
where we used that  $A^i_a=  \frac{\tilde{A}^i_a}{\gamma}$,  $
\Xi^a_i = P^a_i + \pi^a_i, $ $P^a_i = 2e e^a_i$,  and $\tilde{N}=\frac{N}{4e}$, $N^a$ the lapse and shift respectively. Here, we define as usual $F_{ab}^{i} = \partial_{a}\tilde{A}_{b}^{i} - \partial_{b}\tilde{A}_{a}^{i} + \tensor{\epsilon}{^i_j_k}\tilde{A}_{a}^{j}\tilde{A}_{b}^{k} $,  and $R_{ab}^{i} = \partial_{a}\omega_{b}^{i} - \partial_{b}\omega_{a}^{i} - \tensor{\epsilon}{^i_j_k}\omega_{a}^{j}\omega_{b}^{k}$, and 
\begin{eqnarray}
\label{co}
			\pi_{i}^{a} &=& \frac{2}{\gamma}\eta^{abc}\left[F_{bci}- \left(\frac{\gamma^2+1}{\gamma^2}\right)\tensor{\epsilon}{_i_j_k}\left(\omega_{b}^{j}+\tilde{A}_{b}^{j}\right)\left(\omega_{c}^{k}+\tilde{A}_{c}^{k}\right)\right], \nonumber \\
	\tilde{\pi}_{i}^{a} &=& 2\left(\frac{\gamma^2+1}{\gamma^2}\right)\eta^{abc}\left[R_{bci}+ \frac{1}{\gamma^2}\tensor{\epsilon}{_i_j_k}\left(\omega_{b}^{j}+\tilde{A}_{b}^{j}\right)\left(\omega_{c}^{k}+\tilde{A}_{c}^{k}\right)\right].
	\end{eqnarray}
Hence, from (\ref{Holst}) and (\ref{co}) we identify the following constraints
\begin{eqnarray}
\label{cons}
\nonumber 
H &=& \frac{1}{\gamma^2} \tensor{\epsilon}{_i^j^k}\left(\Xi_{j}^{a}-\pi_{j}^{a}\right) \left( \Xi_{k}^{b}-\pi_{k}^{b}\right) \Big[F^i_{ab} + \left(\gamma^2+1\right)R^{i}_{ab} \Big] \approx 0,\\ \nonumber
H_{b} &:&\frac{1}{\gamma}\Xi_{i}^{a}F_{ab}^{i} \approx 0, \nonumber \\
			G_{i} &:& \partial_{a}\Xi_{i}^{a} + \tensor{\epsilon}{_i_j^k}\tilde{A}_{a}^{j}\Xi_{k}^{a} \approx 0,, \nonumber \\
			S_{i} &:& \partial_{a}\Xi_{i}^{a} - \tensor{\epsilon}{_i_j^k}\omega_{a}^{j}\Xi_{k}^{a} \approx 0,\nonumber \\
			D_{i} &:& \partial_{a}\tilde{\pi}_{i}^{a} + \tensor{\epsilon}{_i_j^k}\tilde{A}_{a}^{j}\tilde{\pi}_{k}^{a}\approx 0, \nonumber \\
		E_{i} &:& \partial_{a}\tilde{\pi}_{i}^{a} - \tensor{\epsilon}{_i_j^k}\omega_{a}^{j}\tilde{\pi}_{k}^{a} \approx 0, \nonumber\\
		\tilde{C}^a_i &:&\tilde{\pi}_{i}^{a} - 2\left(\frac{\gamma^2+1}{\gamma^2}\right)\eta^{abc}\left[R_{bci}+ \frac{1}{\gamma^2}\tensor{\epsilon}{_i_j_k}\left(\omega_{b}^{j}+\tilde{A}_{b}^{j}\right)\left(\omega_{c}^{k}+\tilde{A}_{c}^{k}\right)\right]\approx 0, 
	\end{eqnarray}
where $H, H_b, G_i$ are the so-called scalar, vector, and Gauss constraints, respectively, and they are first-class constraints. On the other hand, the constraints  $S_i, D_i, E_i, \tilde{C}^a_i$ are new constraints, and they are of second class. Now, in order to obtain a clossed algebra, the Gauss constrain must to be modified, say, $G_i \rightarrow G_i - E_i$, hence the new Gauss constraint takes the form $G_i= \partial_{a}\Xi_{i}^{a} + \tensor{\epsilon}{_i_j^k}\tilde{A}_{a}^{j}\Xi_{k}^{a} - \partial_{a}\tilde{\pi}_{i}^{a}  + \tensor{\epsilon}{_i_j^k}\omega_{a}^{j}\tilde{\pi}_{k}^{a}$. Now,  the canonical pairs are given by $(\tilde{A}^i_a, \Xi ^a_i)$ and  $(\omega^i _a, \tilde{\pi}^a_i)$, in this manner,  the algebra between the constraints is given by 
\begin{eqnarray}
\nonumber 
			\{G_{i}(\vec{x}), D_{j}(\vec{y})\} &=& \tensor{\epsilon}{_i_j^k}D_k(\vec{x})\delta^3(\vec{x}-\vec{y}),    \\ \nonumber 
		\{G_{i}(\vec{x}), E_{j}(\vec{y})\} &=& \tensor{\epsilon}{_i_j^k}E_k(\vec{x})\delta^3(\vec{x}-\vec{y}),  \\ \nonumber 
			\{G_{i}(\vec{x}), G_{j}(\vec{y})\} &=& \tensor{\epsilon}{_i_j^k}G_k(\vec{x})\delta^3(\vec{x}-\vec{y}),  \\ \nonumber 
			\{D_{i}(\vec{x}), D_{j}(\vec{y})\} &=& \tensor{\epsilon}{_i_j^k}D_k(\vec{x})\delta^3(\vec{x}-\vec{y}),  \\ \nonumber 
			\{E_{i}(\vec{x}), E_{j}(\vec{y})\} &=& - \tensor{\epsilon}{_i_j^k}E_k(\vec{x})\delta^3(\vec{x}-\vec{y}), \\ \nonumber 
	\{D_{i}(\vec{x}), E_{j}(\vec{y})\} &=& \tensor{\epsilon}{_i_j^k}\tilde{\pi}_k^{a}(\vec{x})\partial_a\delta^3(\vec{x}-\vec{y}) + \tensor{\epsilon}{_i_k_l}\tensor{\epsilon}{_j^k^m}\tilde{A}_{a}^{l}(\vec{x})\tilde{\pi}_{m}^{a}(\vec{x})\delta^3(\vec{x}-\vec{y}), \\ \nonumber 
			\{G_{i}(\vec{x}), \tilde{C}_{j}^{a}(\vec{y})\} &=& \tensor{\epsilon}{_i_j^k}\tilde{C}_k^{a}(\vec{x})\delta^3(\vec{x}-\vec{y}), \\ \nonumber 
    \{ \tilde{C}^{a}_{\; k}(\vec x), \tilde{C}^{b}_{\; l}(\vec y) \} &=& 0,\\ \nonumber 
    			\{H[\tilde{N}],H[\tilde{M}]\}
						&=& -\frac{4}{\gamma^2}(\tilde{N}\partial_{a}\tilde{M}-\tilde{M}\partial_{a}\tilde{N})\left(\Xi^{ai}-\pi^{ai}\right)\left(\Xi_{i}^{b}-\pi_{i}^{b}\right)\Biggl\{ H_{b}\\ 
			&-& \left(\frac{\gamma^2+1}{\gamma}\right)\left(\Xi_{j}^{c}-\pi_{j}^{c}\right) R_{bc}^j \Biggr\}, 
			\end{eqnarray}	
however, $\Xi^a_i = P^a_i + \pi^a_i, $ and the Bianchi identity implies that $P^a_i R^i_{ab}=0$ \cite{1}, hence,  the last term of the bracket between the scalar constraint vanishes, and the algebra is closed. We have added an appendix developing the computation of the bracket of $H$ itself. Furtheremore, the counting of degrees of freedom is perfomed as follow: there are 18 canonical variables $(\tilde{A}^i_a, \Xi ^a_i), (\omega^i _a, \tilde{\pi}^a_i)$, seven first-class constraints $(H, H_b, G_i)$ and 18 second class constraints $S_i, D_i, E_i, \tilde{C}^a_i $ thus, the counting of physical degrees of freedom is $DOF=\frac{1}{2}(18- 2(7)-18)=2$, as expected, because we just coupled topological theories to gravity. However, a dependence of the $\gamma$  parameter remains, through the $\pi^a_i$ and $\tilde{\pi}^a_i$ terms. \\
Furthermore, we can observe the contribution of the topological terms to gravity. In fact, if  $\pi^a_i$ is removed,  and $\tilde{\pi}^a_i\approx 0$ (this is because $\omega_a^i$ remains as a canonical variable), then all is reduced to the standard Hamiltonian structure of Holst action \cite{Kari}, say 
\begin{eqnarray}
\label{cons2}
\nonumber 
H &=& \frac{1}{\gamma^2} \tensor{\epsilon}{_i^j^k}P_{j}^{a} P_{k}^{b} \Big[F^i_{ab} + \left(\gamma^2+1\right)R^{i}_{ab} \Big] \approx 0,\\ \nonumber
H_{b} &:&\frac{1}{\gamma}P_{i}^{a}F_{ab}^{i} \approx 0,  \\ \nonumber 
			G_{i} &:& \partial_{a}P_{i}^{a} + \tensor{\epsilon}{_i_j^k}\tilde{A}_{a}^{j}P_{k}^{a} \approx 0, \\ \nonumber 
			S_{i} &:& \partial_{a}P_{i}^{a} - \tensor{\epsilon}{_i_j^k}\omega_{a}^{j}P_{k}^{a} \approx 0, \\  \nonumber 
			D^a_i &:&\tilde{\pi}^a_i\approx 0, \\ 
			\alpha^{ab}&:& \epsilon_{q}{^{jk}}\bigg[ P^{cq}P^a_j \partial_a P^b_k + P^{bq}P^a_j \partial_a P^c_k  - P^a_j \omega^l_a \big[ P^{bi}\epsilon^{q}{_{li}} P^c_l+ P^{ci}\epsilon^{q}{_{li}} P^b_l \big]\bigg]\approx 0, 
			\end{eqnarray}
where the canonical variables are $(\tilde{A}^i_a, P^a_i)$ and $(\omega_a^i, \tilde{\pi}^a_i)$. Hence,  $H, H_b, G_i$ are first-class and $S_i,  D^i_a, \alpha^{ab} $ are second class; $ \alpha^{ab}$ is obtained from concistency of $D_i^a$. From the comparison of (\ref{cons}) and (\ref{cons2}), we can appreciate the difference when the topological invariants are added.  \\
On the other hand, it is worth commenting that if we take $\gamma= \pm i$, then (\ref{Holst}) reproduces the self-dual gravity plus topological terms reported in \cite{13}. In fact, we can observe that $\tilde{\pi}_{i}^{a}, S_i, D_i, E_i, \tilde{C}^a_i $ vanishes, and $\pi^a_k= -2i\eta^{abc}F_{bck}$ with $\tilde{A}^i_a$ being the Ashtekar variables. The action is reduced to 
\begin{eqnarray}
\label{Holst2}
			S &=& \int dx^4 \, \Bigl\{ \frac{ -\Xi_{j}}{i}^{a} \dot{\tilde A}^i_a+ \tilde{\pi}_{i}^{a} \dot{\omega}^a_i  -\tilde{N}\tensor{\epsilon}{_i^j^k}\left(\Xi_{j}^{a}-\pi_{j}^{a}\right)\left(\Xi_{k}^{b}-\pi_{k}^{b}\right)\left[F_{ab}^{i}\right]\\ \nonumber 
			&- &N^{b}\Xi_{i}^{a}F_{ab}^{i} - A_{0}^{i} \left(\partial_{a}\Xi_{i}^{a} + \tensor{\epsilon}{_i_j^k}\tilde{A}_{a}^{j}\Xi_{k}^{a} \right)\Bigl\},  
\end{eqnarray}
where $\tilde N, N^a $ and $\Xi_{k}^{a}$ are defined avobe, and  we have used that $\left[\Xi_{k}^{a}-\pi^a_k \right]F_{ab}^k=\Xi_{k}^{a}F_{ab}^k$, because  $F_{abk}=\frac{i}{4}\eta_{abc}\pi^c_k$ and $\pi^a_i \pi^{bi}\eta_{abc}=0$. Now the constraints are given by 
\begin{eqnarray}
\nonumber
\tensor{\epsilon}{_i^j^k}\left(\Xi_{j}^{a}-\pi_{j}^{a}\right)\left(\Xi_{k}^{b}-\pi_{k}^{b}\right)\left[F_{ab}^{i}\right]&\approx& 0, \\ \nonumber 
\Xi_{i}^{a}F_{ab}^{i} &\approx& 0, \\ 
\partial_{a}\Xi_{i}^{a} + \tensor{\epsilon}{_i_j^k}\tilde{A}_{a}^{j}\Xi_{k}^{a}  &\approx& 0.
\end{eqnarray}
Furthermore, because the election $\gamma= \pm i$ yields complex gravity, it is necessary to add the so-called reality conditions. However, we can observe that the only modification to the Ashtekar constraints is in the momenta, given by the term $\pi^a_k= -2i\eta^{abc}F_{bck}$, so the framework reported in \cite{Ver} will work with our results. In fact, reality constraints can be added, thereby extending the phase space and introducing them as second-class constraints. Thus, the counting of physical degrees of freedom and the gauge symmetry will not be modified.     \\ 
On the other hand, if $\gamma= \pm 1$,  then in (\ref{Holst}) we obtain the Barbero formulation of the Holst action plus $PE$ invariants, a result not reported in the literature. In addition, with $\gamma= \pm 1$, the constraints $ S_i, D_i, E_i, \tilde{C}^a_i$ do not vanish. Then they contribute to the system's canonical structure. In general, those constraints are new for arbitrary $\gamma$ and have not been reported in the literature. \\ 
\section{Conclusions}
In this paper, a canonical description of the $PE$ invariants with a $BI$ parameter has been developed. The $BI$ parameter contributes to the theory's canonical structure through the constraints and the algebra between them. We observed that if   $\gamma=i$, the self-dual formulation of those invariants is recovered. On the other hand, if  $\gamma$ is real, we report the Barbero formulation of these invariants; this result is not reported in the literature. The reducibility conditions between the constraints remain, for arbitrary $\gamma$, and the counting of physical degrees of freedom was performed; the theory is devoid of them, as expected.
Furthermore, we coupled the $PE$ results to the Holst action. We could observe that the $BI$ parameter contributes to the canonical structure. In particular, if $\gamma=i$, the self-dual formulation of the Holst action plus topological terms is recovered. Moreover, if $\gamma$ is real, then the Barbero formulation of the Holst action plus topological invariants is reported. In this respect, new second-class constraints arise in the set, and the complete set of constraints is no longer polynomial; as expected, the nonlinearity of the gravitational field and the topological invariants are manifested in the constraints. \\
We finish with some comments. We observe that the algebra of the second-class constraints does not take an easy form, and that the Dirac brackets are difficult to compute. In this sense, we can use alternative frameworks to construct the fundamental brackets for an arbitrary $\gamma$. In this regard,   in the paper \cite{18}, by using the Faddeev–Jackiw scheme, the fundamental brackets and the quantum states of the $PE$ were reported. In that paper, the Lorentz connexion was worked as a dynamical variable without a $BI$ parameter.  Thus, in this paper, we present tools for analyzing whether the $BI$ parameter contributes to or modifies the quantum structure and the states reported in \cite{18}, as well as the coupling between the Holst action and the invariants, thereby advancing the debate over the $BI$ controversy. All these new ideas are being developed and will be the subject of forthcoming work \cite{19}.    \\
\\
\section{Appendix}
In this appendix, we develop the Poisson bracket between the scalar constraint. We start with the scalar constraint given by 
				\begin{align}
			\label{eq:H}
			\begin{split}
				H &= -\frac{1}{\gamma^2} \tensor{\epsilon}{_i^j^k}\left(\Xi_{j}^{a}-\pi_{j}^{a}\right)\left(\Xi_{k}^{b}-\pi_{k}^{b}\right)\bigg[\partial_{a}\tilde{A}_{b}^{i}-\partial_{b}\tilde{A}_{a}^{i} + \tensor{\epsilon}{^i_l_m}\tilde{A}_{a}^{l}\tilde{A}_{b}^{m}\\
				&+ \left(\gamma^2+1\right)\left(\partial_{a}\omega_{b}^{i}-\partial_{b}\omega_{a}^{i} - \tensor{\epsilon}{^i_l_m}\omega_{a}^{l}\omega_{b}^{m}\right) \bigg] \approx 0,
			\end{split}
		\end{align}
		let 
		\begin{equation}
			U_{ab}^{i} = \partial_{a}\tilde{A}_{b}^{i}-\partial_{b}\tilde{A}_{a}^{i} + \tensor{\epsilon}{^i_{l}_{m}}\tilde{A}_{a}^{l}\tilde{A}_{b}^{m} + \left(\gamma^2+1\right)\left(\partial_{a}\omega_{b}^{i}-\partial_{b}\omega_{a}^{i} - \tensor{\epsilon}{^i_{l}_{m}}\omega_{a}^{l}\omega_{b}^{m}\right),
		\end{equation}
		then we can rewrite (\ref{eq:H}) as
		\begin{equation}
			H = - \frac{1}{\gamma^2}\tensor{\epsilon}{_i^j^k}\left(\Xi_{j}^{a}-\pi_{j}^{a}\right)\left(\Xi_{k}^{b}-\pi_{k}^{b}\right)U_{ab}^{i}.
		\end{equation}
		Using the fundamental Poisson brackets,
			\begin{align}
				\begin{split}
					&\{\tilde{A}_{a}^{i}(\vec{x}),\Xi_{j}^{b}(\vec{y})\} = \gamma\delta_{j}^{i}\delta_{a}^{b}\delta^{3}(\vec{x}-\vec{y}),\\
					&\{\omega_{a}^{0i}(\vec{x}),\tilde{\pi}_{j}^{b}(\vec{y})\} = \delta_{j}^{i}\delta_{a}^{b}\delta^{3}(\vec{x}-\vec{y}),
				\end{split}
			\end{align}	
		we compute the Poisson bracket for the scalar constraint:
		\begin{align*}
			\begin{split}
				&\gamma^4 \{H[\tilde{N}],H[\tilde{M}]\}\\
			    = &\tensor{\epsilon}{_i^j^k}\tensor{\epsilon}{_l^m^n}\Big[\{\left(\Xi_{j}^{a}-\pi_{j}^{a}\right)\left(\Xi_{k}^{b}-\pi_{k}^{b}\right), \left(\Xi_{m}^{c}-\pi_{m}^{c}\right)\left(\Xi_{n}^{d}-\pi_{n}^{d}\right)\}U_{ab}^{i}(\vec{x})U_{cd}^{l}(\vec{y})\\
			    +&\left(\Xi_{j}^{a}-\pi_{j}^{a}\right)\left(\Xi_{k}^{b}-\pi_{k}^{b}\right)\{U_{ab}^{i}(\vec{x}),\Xi_{m}^{c}(\vec{y})\}\left(\Xi_{n}^{d}-\pi_{n}^{d}\right)U_{cd}^{l}(\vec{y})\\
			    +&\left(\Xi_{j}^{a}-\pi_{j}^{a}\right)\left(\Xi_{k}^{b}-\pi_{k}^{b}\right)\{U_{ab}^{i}(\vec{x}),\Xi_{n}^{d}(\vec{y})\}\left(\Xi_{m}^{c}-\pi_{m}^{c}\right)U_{cd}^{l}(\vec{y})\\
			    +&U_{ab}^{i}(\vec{x})\left(\Xi_{j}^{a}-\pi_{j}^{a}\right)\{\Xi_{k}^{b}(\vec{x}),U_{cd}^{l}(\vec{y})\}\left(\Xi_{m}^{c}-\pi_{m}^{c}\right)\left(\Xi_{n}^{d}-\pi_{n}^{d}\right)\\
			    +&U_{ab}^{i}(\vec{x})\left(\Xi_{k}^{b}-\pi_{k}^{b}\right)\{\Xi_{j}^{a}(\vec{x}),U_{cd}^{l}(\vec{y})\}\left(\Xi_{m}^{c}-\pi_{m}^{c}\right)\left(\Xi_{n}^{d}-\pi_{n}^{d}\right)\\
			    +&\left(\Xi_{j}^{a}-\pi_{j}^{a}\right)\left(\Xi_{k}^{b}-\pi_{k}^{b}\right)\{U_{ab}^{i}(\vec{x}),U_{cd}^{l}(\vec{y})\} \left(\Xi_{m}^{c}-\pi_{m}^{c}\right)\left(\Xi_{n}^{d}-\pi_{n}^{d}\right)\Big].
			\end{split}
		\end{align*}
		Now,
			\begin{equation}
				\{\left(\Xi_{j}^{a}-\pi_{j}^{a}\right)\left(\Xi_{k}^{b}-\pi_{k}^{b}\right), \left(\Xi_{m}^{c}-\pi_{m}^{c}\right)\left(\Xi_{n}^{d}-\pi_{n}^{d}\right)\} = 0,
			\end{equation}
			\begin{align}
				\begin{split}
					\{U_{ab}^{i}(\vec{x}),\Xi_{m}^{c}(\vec{y})\}
					&=
					-\gamma\delta_{m}^{i}\left[\delta_{b}^{c}\partial_{a}\left(\delta^{3}(\vec{x}-\vec{y})\right) - \delta_{a}^{c}\partial_{b}\left(\delta^{3}(\vec{x}-\vec{y})\right)\right]\\
					&+ \tensor{\epsilon}{^i_{j'}_m}\left(\delta_{b}^{c}A_{a}^{j'}(\vec{x}) - \delta_{a}^{c}A_{b}^{j'}(\vec{x})\right)\delta^{3}(\vec{x}-\vec{y}),
				\end{split}
			\end{align}
		and
			\begin{equation}
				\{U_{ab}^{i}(\vec{x}),U_{cd}^{l}(\vec{y})\} = 0,
			\end{equation}		
		thus
		\begin{align*}
			\begin{split}
				&-\gamma^3 \{H[\tilde{N}],H[\tilde{M}]\}\\
				=
				&-2\tensor{\epsilon}{_i^j^k}\tensor{\epsilon}{_l^m^n}\left(\Xi_{j}^{a}-\pi_{j}^{a}\right)\left(\Xi_{k}^{b}-\pi_{k}^{b}\right)\left(\Xi_{m}^{c}-\pi_{m}^{c}\right)\delta^{3}(\vec{x}-\vec{y})\\
				*&\delta_{n}^{i}\left(\delta_{b}^{d}M\partial_{a}N-\delta_{a}^{d}M\partial_{b}N\right)\bigg[\partial_{c}\tilde{A}_{d}^{l}-\partial_{d}\tilde{A}_{c}^{l} + \tensor{\epsilon}{^l_{p}_{q}}\tilde{A}_{c}^{p}\tilde{A}_{d}^{q} + \left(\gamma^2+1\right)\left(\partial_{c}\omega_{d}^{l}-\partial_{d}\omega_{c}^{l} - \tensor{\epsilon}{^l_{p}_{q}}\omega_{c}^{p}\omega_{d}^{q}\right)\bigg]\\    
				+&2\tensor{\epsilon}{_i^j^k}\tensor{\epsilon}{_l^m^n}\left(\Xi_{j}^{a}-\pi_{j}^{a}\right)\left(\Xi_{k}^{b}-\pi_{k}^{b}\right)\left(\Xi_{m}^{c}-\pi_{m}^{c}\right)\delta^{3}(\vec{x}-\vec{y})\\
				*&\delta_{n}^{i}\left(\delta_{b}^{d}N\partial_{a}M-\delta_{a}^{d}N\partial_{b}M\right)\bigg[\partial_{c}\tilde{A}_{d}^{l}-\partial_{d}\tilde{A}_{c}^{l} + \tensor{\epsilon}{^l_{p}_{q}}\tilde{A}_{c}^{p}\tilde{A}_{d}^{q} + \left(\gamma^2+1\right)\left(\partial_{c}\omega_{d}^{l}-\partial_{d}\omega_{c}^{l} - \tensor{\epsilon}{^l_{p}_{q}}\omega_{c}^{p}\omega_{d}^{q}\right)\bigg]\\
				=&-4\delta^{jl}\left(\Xi_{j}^{a}-\pi_{j}^{a}\right)\left(\Xi_{l}^{b}-\pi_{l}^{b}\right)\left(\Xi_{k}^{c}-\pi_{k}^{c}\right)\left(M\partial_{a}N-M\partial_{a}N\right)\delta^{3}(\vec{x}-\vec{y})\\
				*&\bigg[\partial_{b}\tilde{A}_{c}^{k}-\partial_{c}\tilde{A}_{b}^{k}+\tensor{\epsilon}{^k_{j'}_{k'}}\tilde{A}_{b}^{j'}\tilde{A}_{c}^{k'}+(\gamma^2+1)\left(\partial_{b}\omega_{c}^{0k}-\partial_{c}\omega_{b}^{0k}-\tensor{\epsilon}{^k_{j'}_{k'}}\omega_{b}^{0j'}\omega_{c}^{0k'}\right)\bigg]\\
				+&4\delta^{kl}\left(\Xi_{j}^{a}-\pi_{j}^{a}\right)\left(\Xi_{l}^{b}-\pi_{l}^{b}\right)\left(\Xi_{k}^{c}-\pi_{k}^{c}\right)\left(M\partial_{a}N-M\partial_{a}N\right)\delta^{3}(\vec{x}-\vec{y})\\
				*&\bigg[\partial_{b}\tilde{A}_{c}^{j}-\partial_{c}\tilde{A}_{b}^{j}+\tensor{\epsilon}{^j_{j'}_{k'}}A_{b}^{j'}A_{c}^{k'}+(\gamma^2+1)\left(\partial_{b}\omega_{c}^{0j}-\partial_{c}\omega_{b}^{0j}-\tensor{\epsilon}{^j_{j'}_{k'}}\omega_{b}^{0j'}\omega_{c}^{0k'}\right)\bigg].
			\end{split}
		\end{align*}
		Therefore
		\begin{align}
			\begin{split}
				&\{H[\tilde{N}],H[\tilde{M}]\}\\
				= &-\frac{4}{\gamma^3}\Big(N(\vec{x})\partial_{a}M(\vec{x})-M(\vec{x})\partial_{a}N(\vec{x})\Big)\\
				&*\bigg[\Big(\Xi_{i}^{a}(\vec{x})-\pi_{i}^{a}(\vec{x})\Big)\Big(\Xi^{bi}(\vec{x})-\pi^{bi}(\vec{x})\Big)\Big(\Xi_{j}^{c}(\vec{x})-\pi_{j}^{c}(\vec{x})\Big)U_{bc}^{j}(\vec{x})\\
				&- \Big(\Xi_{j}^{a}(\vec{x})-\pi_{j}^{a}(\vec{x})\Big)\Big(\Xi^{bi}(\vec{x})-\pi^{bi}(\vec{x})\Big)\Big(\Xi_{i}^{c}(\vec{x})-\pi_{i}^{c}(\vec{x})\Big)U_{bc}^{j}(\vec{x})\bigg].
			\end{split}
		\end{align}
		By the antisymmetry of $U_{bc}^{j}$, the second term is zero, so
		\begin{eqnarray}
		\nonumber 
				\{H[\tilde{N}],H[\tilde{M}]\} &=& 
				-\frac{4}{\gamma^3}\Big(\tilde{N}\partial_{a}\tilde{M}-\tilde{M}\partial_{a}\tilde{N}\Big)\left(\Xi_{i}^{a}-\pi_{i}^{a}\right)\left(\Xi^{bi}-\pi^{bi}\right)\left(\Xi_{j}^{c}-\pi_{j}^{c}\right)U_{bc}^{j}\\\nonumber 
				&=& -\frac{4}{\gamma^2}(\tilde{N}\partial_{a}\tilde{M}-\tilde{M}\partial_{a}\tilde{N})\left(\Xi^{ai}-\pi^{ai}\right)\left(\Xi_{i}^{b}-\pi_{i}^{b}\right)\Biggl\{ H_{b}\\	
				&-& \left(\frac{\gamma^2+1}{\gamma}\right)\left(\Xi_{j}^{c}-\pi_{j}^{c}\right)\Big[\partial_{b}\omega_{c}^{j}-\partial_{c}\omega_{b}^{j}-\tensor{\epsilon}{^j_k_l}\omega_{b}^{k} \omega_{c}^{l}  \Big] \Biggl\} 
		\end{eqnarray}
we can observe that the presence of the $\gamma$ parameter is given through $\pi^a_i$. It is worth noting that the scalar constraint can be written in a new fashion, similar to that expressed in Loop gravity. In fact, by using the change of variable, 
$\omega^{0i}_\alpha= A^i_{\alpha}- \frac{\omega^i_\alpha}{\gamma} $ in the definition of  $F_{ab}^{i} = \partial_{a}\tilde{A}_{b}^{i} - \partial_{b}\tilde{A}_{a}^{i} + \tensor{\epsilon}{^i_j_k}\tilde{A}_{a}^{j}\tilde{A}_{b}^{k} $, we obtain 
\begin{eqnarray}
\label{F}
F_{ab}^i= - R_{ab}^i + \gamma (D_a \omega^{0i}_b -D_b \omega^{0i}_a) + \gamma^2\epsilon^{i}{_{jk}} \omega^{0i}_a\omega^{0j}_b, 
\end{eqnarray}
hence, by considering (\ref{F}) in the scalar constraint, we get 
\begin{eqnarray}
\label{H}
H=  \tensor{\epsilon}{_i^j^k}\left(\Xi_{j}^{a}-\pi_{j}^{a}\right)\left(\Xi_{k}^{b}-\pi_{k}^{b}\right)\bigg[F^i_{ab}+2(\gamma^2 +1) \omega^{0i}_{[a}\omega^{0j}{_{b]}}\bigg], 
\end{eqnarray}
where the constraint $G_i$ was taken into account, and this form is that reported in the literature \cite{Ash}, plus the contribution of the $\pi^a_i$ terms. 

\end{document}